\documentclass[11pt]{article}

\usepackage{amsmath}
\usepackage{amssymb}

\usepackage{graphicx}
\usepackage{dcolumn}
\usepackage{bm}

\usepackage{color}
\usepackage[dvipsnames]{xcolor}
\usepackage{soul,cancel}
\usepackage[T1]{fontenc}

\usepackage{lipsum}
\usepackage{amsfonts}
\usepackage{graphicx}
\usepackage{epstopdf}
\usepackage{algorithmic}
\ifpdf
  \DeclareGraphicsExtensions{.eps,.pdf,.png,.jpg}
\else
  \DeclareGraphicsExtensions{.eps}
\fi

\usepackage{ulem}					
\usepackage{amsmath,amssymb,hyperref,graphicx}

\definecolor{MyBlue}{rgb}  {0.1,0.1,0.9}
\definecolor{MyRed}{rgb}   {0.9,0.1,0.1}
\definecolor{MyGreen}{rgb} {0.05,0.4,0.05}
\definecolor{burntorange}{rgb}{0.8, 0.33, 0.0}

\definecolor{fuchsia}{rgb}{1.0, 0.0, 1.0}

\newcommand{\BBC}{\mathbb{C}}      
\newcommand{\BN}{\mathbb{N}}      

\newcommand{\BJ}{\mathbf{J}}
\newcommand{\BA}{\mathbf{A}}
\newcommand{\BB}{\mathbf{B}}
\newcommand{\BC}{\mathbf{C}}
\newcommand{\BD}{\mathbf{D}}

\newcommand{\lapl}{\Delta}				
\renewcommand{\phi}{\varphi}
\DeclareMathOperator{\tr}{tr}			
\def\d{\,{\rm d}}									

\usepackage[colorinlistoftodos,textsize=tiny,obeyFinal]{todonotes}

\title{Pattern formation in reaction-diffusion systems with piece-wise kinetic modulation: an example study of heterogeneous kinetics.}

\author{Michal Koz{\'a}k $^1$, Eamonn A. Gaffney $^2$, V{\'a}clav Klika $^{1,*}$}%

\date{\today}

\begin{document}

\maketitle

\vspace*{3mm}\noindent
$^1$ Department of Mathematics, FNSPE, Czech Technical University in Prague, Prague, Czech Republic\\
$^2$ Wolfson Centre for Mathematical Biology, Mathematical Institute, University of Oxford, U.K.\\
$^*$ Corresponding author: vaclav.klika@fjfi.cvut.cz

   %

\subsection*{Abstract}
  The study of pattern emergence together with exploration of the exemplar Turing model is enjoying a renaissance both from theoretical and experimental perspective.
  Here, we implement a stability analysis of spatially dependent reaction kinetics by exploring the effect of a jump discontinuity within piece-wise constant kinetic parameters, using various methods to identify and confirm the diffusion-driven instability conditions. Essentially, the presence of stability or instability in Turing models is a local property for piece-wise constant kinetic parameters and, as such, may be analysed locally. In particular, a local assessment of whether parameters are within the Turing space provides a strong indication that for a large enough region with these parameters, an instability can be excited.

\section{Introduction}\label{sec_intro}

Understanding pattern formation is one of the major issues not only in developmental biology but across many different disciplines. 
A seminal mechanism for self-organisation emergence was proposed by Alan Turing in 1952 \cite{Turing1952}. He considered two biochemicals (the so-called morphogens) which diffuse and interact with each other via reaction terms and demonstrated that a small fluctuation of their concentrations around a steady state can be heterogeneously amplified for suitable reaction kinetics coupled with diffusion. Therefore spatially non-homogeneous steady states, that is patterns, can emerge. Mathematically, this symmetry breaking mechanism was described as a diffusion-driven instability (DDI) of the steady state in a reaction-diffusion system (RD system). Comparing theoretical predictions and real patterns seen in nature, it has been reported that Turing models may indeed feature in, for example, the formation of wild cat skin patterns \cite{Maini2006}, tumour vascularisation \cite{Chaplain2001}, hair follicle localisation \cite{Klika2012, GloverPlos2017}, pigmentation of zebrafish \cite{Nakamasu2009}, rugae formation \cite{Pantalacci2008} and digit patterning \cite{Sheth2012}.

For many decades, one of the biggest issues for the plausibility of a Turing system in biological development was the lack of identification of morphogens and confirming the molecular details match those required for patterning, though recent studies show extensive promise \cite{Kondo2010, Economou2012, Raspopovic2014}. Furthermore, there is in addition the suggestion that the combination of Wolpert's positional information hypothesis \cite{Wolpert1969, Green2015} with Turing's mechanism may increase the applicability of both \cite{Miura2013, Economou2013}.

Many Turing systems have been proposed and analysed, both analytically and using numerical experiments. The majority of these studies consider only two species. The exemplar models are the Gierer-Meinhardt model \cite{Gierer1972} describing the growth of the hydra; the Schnakenberg model \cite{Schnakenberg1979} describing a chemical reaction exhibiting limit-cycle behaviour or the Thomas model \cite{Thomas1975} describing chemical reaction of oxygen and uric acid in presence of the enzyme uricase; see \cite{Murray2003} for a review of these models. A standard question is to find the Turing space, that is the set of parameters for which Turing's diffusion-driven instability occurs, whereby the system's steady state is stable to homogeneous perturbations but unstable to heterogeneous perturbations assuming the domain size is sufficiently large to support the growth of the unstable perturbations.


The resulting patterns of the Turing model with constant coefficients are typically highly periodic and hence fail to capture key features of many real systems, where the patterns significantly change in space. The reason is intuitive -- the model is too simplistic. The addition of spatial dependency in model parameters is a natural way to extend the Turing model and probably an intuitive way of potentially generating wavelength variation \cite{miguez2006effect}. The Gierer-Meinhardt model \cite{Gierer1972}, for example, is originally based on a source with a gradient; even Turing in his paper \cite{Turing1952} discussed that ``Most of an organism, most of the time, is developing from one pattern into another, rather than from homogeneity into a pattern'', suggesting significant spatial dependency may be initially present before the impact of the Turing instability.

There are many examples of spatial irregularity in patterns where  wavelength variation is manifest, for example a distribution of mouse or cat whiskers \cite{Painter2012}, alternating thin and thick stripes of Lionfish \cite{Page2005} and emergence of fingers \cite{Economou2013}. In these cases heterogeneous pattern modulation is a crucial feature of the self-organisation and, as a result, we would like to capture this phenomena in a modelling framework for the emergence of self-organisation. Similarly, the heterogeneity of the environment in landscape ecology \cite{pickett1995} or the influence of geometric confinement of human embryonic stem cells \cite{warmflash2014} illustrate important and heterogeneous aspects of patterning systems. Such ubiquitous examples emphasise the critical importance of studying the impact of heterogeneity and its effects on the properties of self-organising systems.

However, the standard procedure of stability analysis is not easily extendable to the case with spatially dependent coefficients as, for example, a homogeneous steady state does not exist in general given spatially dependent kinetics. Probably the most well understood effect of heterogeneity comes from the shadow limit in Gierer-Meinhardt kinetics using spike solutions \cite{iron2001,wei2009,Ward2002} but this requires the applicability of the shadow limit. Further the case of spatially dependent diffusion coefficient was analysed in \cite{Benson1993} with a step function representing the dependency. Heterogeneity in the reaction kinetics was analysed numerically \cite{Page2005} and limited analytical progress in stability analysis has been established with spatial dependency in the kinetics. Examples include asymptotically small, spatially dependent, linear gradients of morphogen source \cite{glimm2009,glimm2012,glimm2017}, a cosine spatial dependence in a coefficient of the kinetics \cite{Mau2012} and a step function, independent of morphogen concentration, added to the kinetics \cite{page2003}.

This article deals with a system with a spatially dependent coefficient in the linear term of the activator kinetics which introduces very different challenges to the above described cases. For analytical convenience, we assume a one dimensional space. We consider the following RD-system:
\begin{equation}\label{sys_RD}
	\begin{aligned}
		\partial_t u &= d_1 \partial_{xx}u + f_1(u,v) + h(x)u, \quad x\in(0,L)\\
		\partial_t v &= d_2 \partial_{xx}v + f_2(u,v), \quad x\in(0,L),
	\end{aligned}
\end{equation}
with Neumann boundary conditions
\begin{equation}\label{BC_N}
	\begin{aligned}
		\frac{\partial u}{\partial n}=0,\quad \mbox{ at } x=0,L \\
		\frac{\partial v}{\partial n}=0,\quad \mbox{ at } x=0,L,
	\end{aligned}
\end{equation}
where $h(x)$ is a step function defined as
\begin{equation}\label{fce_h}
	h(x)=\begin{cases}
		0 & x \in [0,\xi),\\
		s & x \in [\xi,L].
	\end{cases}
\end{equation}

First for reasons that shall become evident, we need to assess what will be denoted as a pattern since the standard definition is not sufficient because inhomogeneity is always present due to the forced jump $s$ at location $\xi$. We illustrate the effect of the step function via  numerical simulations, considering system \eqref{sys_RD} with Schnakenberg kinetics
\begin{equation}\label{fce_schn}
	\begin{aligned}
		f(u,v) &= a-u +u^2v,\\
		g(u,v) &= b-u^2v,
	\end{aligned}
\end{equation}
where $a$, $b$ are positive parameters. Let the step, with a size $s=0.5$, be located in the middle of the domain $\xi=L/2$ and consider the remaining parameters $d_1, d_2, a, b$ to be outside ($b\in \{0.01,0.25\}$) or inside ($b \in \{1,2\}$) the Turing space for $s=0$.

\begin{figure}
	\centering
	\includegraphics[width=0.4\textwidth]{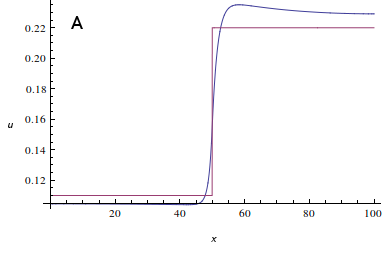}
	\includegraphics[width=0.4\textwidth]{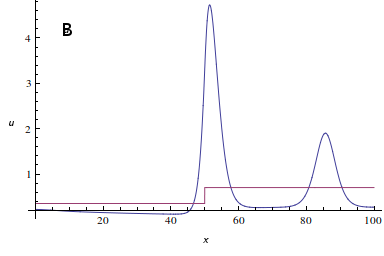}
	\includegraphics[width=0.4\textwidth]{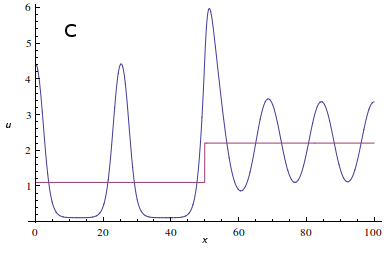}
	\includegraphics[width=0.4\textwidth]{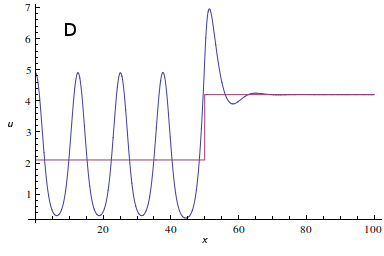}

	\caption{\label{fig_res1} Plot of the large-time (close to steady state) activator concentration $u(x)$ from simulations of the system \eqref{sys_RD} with Schnakenberg kinetics \eqref{fce_schn}, on the domain $[0,100]$ with zero-flux boundary conditions, and parameters: $d_1=1$, $d_2=100$, $s=0.5$, $a=0.1$ and A) $b=0.01$, B) $b=0.25$ (both outside the Turing space for $s=0$), C) $b=1$, D) $b=2$ (both inside the Turing space for $s=0$). These have been solved by Mathematica (for more details see below). Note the different vertical scales in the above plots.}
\end{figure}

As we can see in Fig. \ref{fig_res1} (blue wavy line), a spatial inhomogeneity occurs in each of the solutions. Note i) the different amplitudes of the pattern in the two parts of the domain (as already observed, e.g., in \cite{page2003}) and ii) the different periods (see Fig. \ref{fig_res1}C). Characterising such patterning behaviour in Fig \ref{fig_res1}c, within the context of emergent Turing self-organisation from a linearised system with spatial heterogeneity in the form of a step function premultiplying a linear term, constitutes the overarching aim of this study.

\begin{figure}
	\centering
	\includegraphics[width=0.4\textwidth]{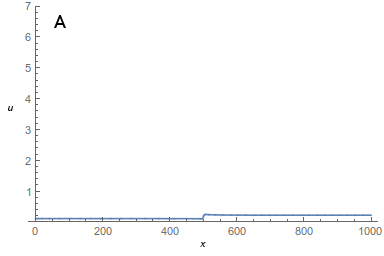}
	\includegraphics[width=0.4\textwidth]{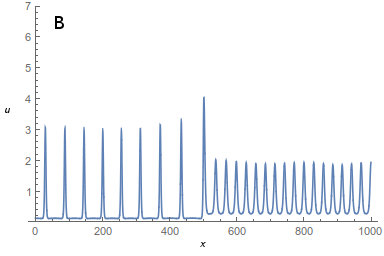}
	\includegraphics[width=0.4\textwidth]{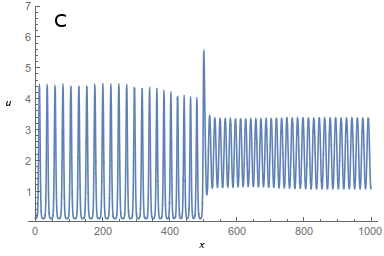}
	\includegraphics[width=0.4\textwidth]{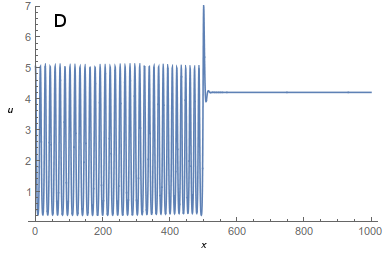}

	\caption{\label{fig_res2} Plot of the large-time (close to steady state) activator concentration $u(x)$ from simulations of the system \eqref{sys_RD} with the same values as in Fig. \ref{fig_res1} apart from $L=1000$. The vertical axis is chosen equal for all plots. This Figure illustrates the effect of larger $L$ which is key for discussion of what a ``pattern'' should mean in systems with heterogeneous kinetics.
	}
\end{figure}

As mentioned above, not all the inhomogenous solutions displayed in Fig. \ref{fig_res1} correspond to a pattern however. To represent genuine self-organisation, rather than being passively slave to the step function, such stationary solutions should have spatial oscillations extending to the domain edge on at least one side of the step even as the domain size is increased, for sufficiently large domain sizes. 
This requirement follows from an observation that a Turing pattern is characterised by a finite number of frequencies that appear in the pattern. Thus an increase in the domain size should result in pattern repetition over the whole domain once a critical domain size is surpassed.
Therefore we plot stationary solutions to the system with the same parameters except a larger domain size, $L=1000$, in Fig. \ref{fig_res2}
. By comparison, we can deduce that the case A is not a pattern as the inhomogeneity is localised only around the point of the step $\xi$ while being of the order of the step $s$. Hence we disregard such cases in the context of pattern. On the contrary, in the other examples the inhomogeneity perseveres on the whole domain; thus such cases are denoted as a pattern.

Hence our detailed objectives are: (i) to determine if a pattern emerges or not; (ii) to undertake a more specific pattern classification examining the parameter spaces for when the system will exhibit each prototype of a stationary solution represented by the plots in Fig. \ref{fig_res1} -- A) no pattern, B) right-sided pattern, C) global pattern, D) left-sided pattern\footnote{Note the discrepancy between plots in Fig. \ref{fig_res1}B and Fig. \ref{fig_res2}B: the former case is denoted as a right-sided pattern whereas the latter patterns globally. The only change is in the value of parameter $L$ and it seems, roughly speaking, $L$ should be large enough and cases where pattern cannot form solely because of domain are generally classified as patterning systems nonetheless.}.

Let us further make the following interesting observation. We consider step functions $\bar{u}(x),\bar{v}(x)$ defined as
\begin{equation}\label{def_PSS}
	\bar{u}(x) = \begin{cases}
		\bar{u}^L & x \in (0,\xi) \\
		\bar{u}^R & x \in (\xi,L)
		\end{cases},
	\quad\quad\quad
	\bar{v}(x) = \begin{cases}
		\bar{v}^L & x \in (0,\xi) \\
		\bar{v}^R & x \in (\xi,L)
		\end{cases},
\end{equation}
where
\begin{equation*}
	\begin{aligned}
		f(\bar{u}^L, \bar{v}^L) &=0 = g(\bar{u}^L, \bar{v}^L), \\
		f(\bar{u}^R, \bar{v}^R) +s\bar{u}^R &=0 = g(\bar{u}^R, \bar{v}^R).
	\end{aligned}
\end{equation*}
Note that in Turing-like problems the existence of a homogeneous steady state is assumed, guaranteeing existence of the step functions \eqref{def_PSS}. The large-time solution for the variable, $u$, seems to be either approaching $\bar{u}(x)$ in the case without pattern, or oscillating around $\bar{u}(x)$ in the case of pattern. This leads us to a hypothesis that the behaviour of the large-time solution could be deduced from attracting properties of the system around $\bar{u}(x)$ and therefore the system characterisations could be similar to the conditions of diffusion driven instability evaluated for the system (with a constant coefficient) considered separately on intervals $(0,\xi)$ and $(\xi,L)$. In particular, such prospects will be considered more precisely and systematically below.

In Section \ref{LinKin.spectral}, we consider system \eqref{sys_RD} with affine kinetics $f,g$ and linearise about the steady state. The system behaves similarly to the original one, but where the propensity to pattern is replaced by the existence of an unbounded solution as time tends to infinity. Hence, we analyse the growth or decay of perturbations in the linearised system, with the implicit assumption that the eventual steady state of the full system will inherit the same spatial frequency of heterogeneity of one of the unstable, diffusivelly driven, growing solutions exhibited by the linearised system. Such implicit assumptions are standard in linear stability analysis \cite{Murray2003}. Furthermore when the linear system exhibits no growing solutions, even for an arbitrarily large domain, the system does not exhibit a Turing instability. Thus, in summary, we analytically solve the stationary problem and proceed to analyse 
the linear stability of the steady state via a spectral approach with the results compared to extensive numerical simulations of the full evolution problem.

We then complement this spectral approach by a boundary layer analysis which further suggests that conditions for one-sided pattern are related to Turing conditions considered separately on each subinterval. In addition, this boundary layer analysis indicates that spatial frequency of emerging pattern is determined independently in the two subintervals while the boundary layer highlights how the patterns match in the interval around the step at $x=\xi$ of the step functions. Finally, the hypothesised local conditions for a Turing instability are verified by a sweep of numerical solutions to the evolution problem with Schnakenberg kinetics before drawing conclusions in Section \ref{sec_disc}.



\section{Analytical approach}

\subsection{Spectral theory} \label{LinKin.spectral}


Let us consider  system \eqref{sys_RD} with general kinetics. The behaviour of such a system is usually inferred from the behaviour of the linearised system, which describes the evolution around the steady state where the influence of linear kinetics exceeds the influence of non-linear parts of kinetics. Hence characterisations via linear systems are generally true at least for small enough initial perturbations of the steady state.

However, as mentioned in Section \ref{sec_intro}, since $s$ is non-zero, we cannot expect the steady state to be homogeneous.
 To identify a steady state solution in the non-linear case requires solving a non-linear elliptic partial differential equation. 
Further, if we consider an expansion of the non-linear kinetics around this non-homogeneous steady state, the resulting coefficients in the linear terms would be strongly spatially dependent which may prevent analytical tractability. 
Hence we focus on the case of linear kinetics where the step occurs only in a single linear kinetic term. Then, generalisation to linear kinetics with jumps in every kinetic term is implemented.


Consider a following RD system with affine kinetics and the step function $h(x)$ defined above \eqref{fce_h}
\begin{equation}\label{sys_RD_lin}
	\begin{aligned}
		\partial_t u &= d_1 \partial_{xx}u + b_{10}+(b_{11}+h(x))u +b_{12}v\\
		\partial_t v &= d_2 \partial_{xx}v + b_{20}+b_{21}u +b_{22}v
	\end{aligned}
	\quad x\in(0,L),
\end{equation}
with Neumann boundary conditions \eqref{BC_N}.

The advantage of considering affine kinetics, even though they are not standard, is that the system then in general describes the dynamics not only around a trivial solution, but also around a non-trivial one. 

\subsubsection{Analytic solution to the stationary problem}

Our first task is to find a stationary solution to the system \eqref{sys_RD_lin} explicitly. The following approach is not novel, moreover the complexity and length of the steady state solution is out of proportion to the usefulness of explicitly stating it and hence we highlight the important features only.

Since $\bar{u}(x)$, $\bar{v}(x)$ defined in \eqref{def_PSS} and $h(x)$ are step functions at $\xi$, it is easier to consider the system~\eqref{sys_RD_lin} restricted to the interval~$(0,\xi)$ and $(\xi,L)$ separately and to use appropriate connecting conditions at $\xi$. From now on the upper index $L$ corresponds to the system on $(0,\xi)$, whilst the index $R$ corresponds to the interval $(\xi,L)$ respectively. For the perturbations
\begin{equation*}
	\begin{aligned}
		\tilde{u}^L &= u^L-\bar{u}^L, & \tilde{v}^L &= v^L-\bar{v}^L,\\
		\tilde{u}^R &= u^R-\bar{u}^R, & \tilde{v}^R &= v^R-\bar{v}^R
	\end{aligned}
\end{equation*}
we obtain an equivalent steady system of the form
\begin{equation}\label{sys_RD_linh}
	\begin{aligned}
		&\begin{aligned}
			0 &= d_1 \partial_{xx}\tilde{u}^L + b_{11}\tilde{u}^L +b_{12}\tilde{v}^L,\\
			0 &= d_2 \partial_{xx}\tilde{v}^L + b_{21}\tilde{u}^L +b_{22}\tilde{v}^L,
		\end{aligned}
			& &\quad \text{in $(0,\xi)$,}\\
		&\begin{aligned}
			0 &= d_1 \partial_{xx}\tilde{u}^R + (b_{11}+s)\tilde{u}^R +b_{12}\tilde{v}^R,\\
			0 &= d_2 \partial_{xx}\tilde{v}^R + b_{21}\tilde{u}^R +b_{22}\tilde{v}^R,
		\end{aligned}
			& &\quad \text{in $(\xi,L)$,}
	\end{aligned}
\end{equation}
with boundary and connecting conditions ($C^1(0,L)$ solutions of the form $u=u^L h(\xi-x)+u^R h(x-\xi)$, $v=v^L h(\xi-x)+v^R h(x-\xi)$)
\begin{equation}\label{sys_spWlin}
	\begin{aligned}
		\frac{\partial \tilde{u}^L}{\partial x}(0) &=\frac{\partial \tilde{v}^L}{\partial x}(0)=0, &
		\frac{\partial \tilde{u}^R}{\partial x}(L) &=\frac{\partial \tilde{v}^R}{\partial x}(L)=0,\\
		\frac{\partial \tilde{u}^L}{\partial x}(\xi) &= \frac{\partial \tilde{u}^R}{\partial x}(\xi), &
		\frac{\partial \tilde{v}^L}{\partial x}(\xi) &= \frac{\partial \tilde{v}^R}{\partial x}(\xi),\\
		\tilde{u}^R(\xi)-\tilde{u}^L(\xi)&= \bar{u}^L-\bar{u}^R, & \tilde{v}^R(\xi)-\tilde{v}^L(\xi)&= \bar{v}^L-\bar{v}^R.
	\end{aligned}
      \end{equation}
Tildes will be omitted henceforth. 

The simple way to solve such systems is to transform the matrix $\BB=\begin{pmatrix}b_{11} & b_{12}\\ b_{21} & b_{22}\end{pmatrix}$ into its Jordan normal form on both subsystems separately, considering the boundary conditions but not the connecting ones, and completing the calculation of the steady state by employing the connecting conditions, as implemented in \cite{page2003}.

An interesting issue is that the resulting form of the solution to each subsystem depends on the sign of $(\tr^2 \BA^{L,R}-4\det\BA^{L,R})$, where $\BA^{L,R}$ denote matrices of coefficients readily written in terms of normalised linear terms for both subsystems, i.e.
\begin{equation*}
	\BA^L = \begin{pmatrix}b_{11}/d_1 & b_{12}/d_1 \\ b_{21}/d_2 & b_{22}/d_2\end{pmatrix}, \quad \BA^R =  \begin{pmatrix}(b_{11}+s)/d_1 & b_{12}/d_1 \\ b_{21}/d_2 & b_{22}/d_2\end{pmatrix}.
\end{equation*}
Particularly, the signs of the following two terms are important
\begin{equation}
	\begin{aligned}
		(d_2b_{11}+d_1b_{22})^2-4d_1d_2\det{\BB},\\
		(d_2(b_{11}+s)+d_1b_{22})^2-4d_1d_2(\det{\BB}+s b_{22}).\\
	\end{aligned}
\end{equation}
These terms are remarkably similar to these appearing on one of the conditions for classical Turing diffusion-driven instability evaluated on both subsystems separately. This observation supports our hypothesis at the end of Section \ref{sec_intro} about the relation between diffusion-driven instability in the studied case and DDI conditions for patterns on $(0,\xi)$ and $(\xi,L)$.


\subsubsection{Linear stability} \label{sect_linStab}

With the steady state of system \eqref{sys_RD_lin}, we can analyse its stability. We focus only on the large-time behaviour of the system, which can be obtained from spectral analysis, assuming that the transient behaviour is not essential as shown for the classical Turing instability \cite{klika2017Chaos}.

Let us denote the (non-constant) steady state as $(\hat{u}, \hat{v})$ and, with a redefinition of $\tilde{u},\tilde{v}$ to the time dependent perturbed solution, $\tilde{u} = u-\hat{u}$, $\tilde{v}=v-\hat{v}$  expand the evolution equations around the steady state to find
\begin{equation} \label{Eq.10}
	\begin{aligned}
		\partial_t \tilde{u} &= d_1 \partial_{xx}\tilde{u} + (b_{11}+h(x))\tilde{u} +b_{12}\tilde{v}\\
		\partial_t \tilde{v} &= d_2 \partial_{xx}\tilde{v} + b_{21}\tilde{u} +b_{22}\tilde{v}.
	\end{aligned}
\end{equation}
Since we consider Neumann boundary conditions, we have a complete orthogonal basis $y_n(x)$, $n \in \BN_0$ of $L^2(0,L)$ and eigenvalues $\kappa_n=(n\pi/L)^2$ for the negative Laplacian (which satisfy $-\lapl y_n = \kappa_n y_n$). Now we rewrite functions $\tilde{u}$ and $\tilde{v}$ in terms of the series
\begin{equation}
	\begin{aligned}
		\tilde{u}(t,x) &= \sum_{n=0}^\infty A_n(t)y_n(x), & \tilde{v}(t,x) &= \sum_{n=0}^\infty B_n(t)y_n(x).
	\end{aligned}
\end{equation}
Thus system \eqref{Eq.10} can be rewritten into the form
\begin{equation} \label{Eq.12}
	\sum_{n=0}^\infty \begin{pmatrix} \partial_t A_n \\ \partial_t B_n\end{pmatrix} y_n(x)+ \BD\begin{pmatrix}A_n \\ B_n\end{pmatrix} \kappa_n y_n(x) - \BJ(x)\begin{pmatrix}A_n \\ B_n\end{pmatrix} y_n(x) = 0,
\end{equation}
where we have introduced a standard notation
\begin{equation}
	\begin{aligned}
		\BD&=\begin{pmatrix}d_1 & 0\\0 & d_2\end{pmatrix} & \BJ(x)&=\begin{pmatrix}J_{11}(x) & J_{12} \\J_{21} & J_{22}\end{pmatrix} = \begin{pmatrix} b_{11}+h(x) & b_{12} \\ b_{21} & b_{22}\end{pmatrix}.
	\end{aligned}
      \end{equation}

Problems of the type \eqref{Eq.12} can be solved using spectral methods, i.e. using expansions in eigenfunctions of the negative Laplacian, see Appendix \ref{sec_lin} for details. However, the system does not decouple into a straightforward collection of coupled ordinary differential equations for amplitudes of the eigenfunctions but rather a truncation has to be employed with detailed calculations presented in Appendix \ref{sec_lin}, with results summarised in Appendix \ref{AppA31}. Note that extension to a general linear case with jumps in every term is straightforward, see Appendix \ref{lin.general}.

Note that the same qualitative behaviour (with the same large time behaviour) is obtained when studying a perturbation of the piece-wise constant solution \eqref{def_PSS} in a generalised function sense, see Appendix \ref{App.delta}, or when instead of linearising the whole problem we linearise just the kinetics, see Appendix \ref{App.LinearisationOfKinetics} (in particular compare Eqns. \eqref{sys_RDL} and \eqref{sys_RD_linFull} and also compare Eqns. \eqref{25half} and \eqref{27}, noting Appendix \ref{lin.general}).

Although we show below that the analysis of Eqn. \eqref{Eq.12} does yields correct results (via comparison to numerical solutions), this linear analysis does not allow the study of the most interesting and pertinent phenomena (being also the main motivation for this study), the influence of heterogeneity on spatial frequency of a pattern and the one-sided patterns.

\subsection{Boundary layer analysis}

To proceed with the analysis for general kinetics and to facilitate a boundary layer analysis we regularise the Heaviside replacing $h(x)$ with
$$ h_\delta(x) = \frac{s}{2}\left[1+ g\left(\frac{x-\xi}{\delta}\right)\right],\quad \mbox{ with } g\in C^\infty(R), \lim_{x\rightarrow \pm \infty} g(x) = \pm 1,~g'\geq 0$$
where one can think of, for example,
$$h_\delta(x)=\frac s 2 \left[1+\tanh\left(\frac{x-\xi}{\delta}\right)\right]$$
and consider  small values of $\delta >0$.  Then the steady state (assuming it exists), $(u_s,v_s)$, satisfies 
\begin{equation}\label{sys_RDL_000}
	\begin{aligned}
		0 &= d_1 \partial_{xx}u_s + f(u_s,v_s)+h_\delta(x) u_s  \\
		0 &= d_2 \partial_{xx}v_s +  g(u_s,v_s)
	\end{aligned}\quad\text{on $(0,L)$.}
\end{equation}

Expanding about the steady state that is not designated to be a pattern, as described earlier, we have 
$$ u = u_s +\tilde{u}, ~~~~v = v_s +\tilde{v},$$ with 
\begin{equation}\label{sys_RDL_00000}
	\begin{aligned}
	\tilde{u}_t &= d_1 \partial_{xx}\tilde{u} + J_{11}(u_s,v_s) \tilde{u}+ J_{12}(u_s,v_s) \tilde{v} +h_\delta(x) \tilde{u} \\
		\tilde{v}_t  &= d_2 \partial_{xx}\tilde{v} +  J_{21}(u_s,v_s) \tilde{u}+  J_{22}(u_s,v_s) \tilde{v}
	\end{aligned}~~~~~~~~~~\quad\text{on $(0,L)$,}
\end{equation}
where ${\bf J}(u_s,v_s)$ is the Jacobian of the kinetics about the steady solution $(u_s,v_s)$. 

We proceed with a boundary layer analysis. Based on the continuity of solution on data we argue that for small enough jump $s$ the steady state solution that does not correspond to a pattern will be approximately   $(\bar{u}^L, \bar{v}^L)$ sufficiently to the left of $x=\xi$ and approximately $(\bar{u}^R, \bar{v}^R)$ sufficiently to the right, where 
\begin{equation*}
	\begin{aligned}
		f(\bar{u}^L, \bar{v}^L) &=0 = g(\bar{u}^L, \bar{v}^L) \\
		f(\bar{u}^R, \bar{v}^R) +s\bar{u}^R &=0 = g(\bar{u}^R, \bar{v}^R).
	\end{aligned}
\end{equation*}
Hence for $x<\xi$, $|x-\xi|\gg \delta$ we anticipate the approximation 
 \begin{equation}\label{sys_RDL_0000000}
	\tilde{u}_t = d_1 \partial_{xx}\tilde{u} + J_{11}(\bar{u}^L, \bar{v}^L)\tilde{u}+ J_{12}(\bar{u}^L, \bar{v}^L) \tilde{v}  , \qquad
		\tilde{v}_t= d_2 \partial_{xx}\tilde{v} +  J_{21}(\bar{u}^L, \bar{v}^L) \tilde{u}+  J_{22}(\bar{u}^L, \bar{v}^L) \tilde{v}
\end{equation} and similarly 
\begin{equation}\label{sys_RDL_000000000}
	\tilde{u}_t = d_1 \partial_{xx}\tilde{u} + J_{11}(\bar{u}^R, \bar{v}^R)\tilde{u}+ J_{12}(\bar{u}^R, \bar{v}^R) \tilde{v}  +s \tilde{u} , \qquad
		\tilde{v}_t  = d_2 \partial_{xx}\tilde{v} +  J_{21}(\bar{u}^R, \bar{v}^R) \tilde{u}+  J_{22}(\bar{u}^R, \bar{v}^R) \tilde{v}
\end{equation}
for $x>\xi$, $|x-\xi|\gg \delta$.
These can be considered as the outer problems for a leading order boundary layer approximation. 

We proceed to consider the prospects of an  internal  boundary layer near  $x\approx \xi$. Indeed noting the form of $h_\delta(x)$, which drives hetergeneous  behaviour near $x=\xi$, one can rescale  the spatial component via $$X=(x-\xi)/\delta.$$ This will lead to 
the absence of a dominant balance   for an inner expansion with $\tilde{u} \sim u_{in}(X,t)+o(1)$, which instead yields
$ d_1 \partial_{XX}u_{in}=0$ at leading order, with an analogous observation for $\tilde{v}  \sim v_{in}(X,t)+o(1)$. The resulting linear solution behaviour is divergent as  $|X|\rightarrow \infty$ 
 unless $u_{in}$ and $v_{in}$ are  constant, indicating no boundary layer, but instead a matching of the left and right outer solutions, via a nominal but constant inner layer solution.  Given the kinetics are assumed to be order unity as $\delta$ is decreased, a  dominant balance at leading order, and thus more complex dynamics,  is only conceivable with  the concomitant temporal rescaling, $T=t/\delta^2$. Then, at leading order 
$$ \partial_T u_{in} = d_1 \partial_{XX}u_{in} , ~~~~~~ \partial_T v_{in} = d_2 \partial_{XX}v_{in} ,$$
with the kinetics subdued by a factor of $\delta^2$. 
The resulting  dynamics is very fast and, more importantly,  
 pure diffusion. Thus it will not drive patterning within the inner region but instead instigate diffusion on a very fast timescale, acting to homogenise across the inner region, whereby for $T\gg 1 $, i.e. $t\gg\delta^2$, one will expect an inner solution which is approximately constant after transients have relaxed even if the initial conditions are highly varying in the vicinity of $x\approx\xi$.
 
 
 Hence,  considering the impact of $\delta \ll 1$, with the limit of zero $\delta$ corresponding to the Heaviside function of interest in the kinetics, the evidence is that the inner solution of a boundary layer analysis  does not induce patterning but has a rather trivial dynamics. Instead,  the behaviour of the outer solutions, i.e. 
 Eqns. (\ref{sys_RDL_0000000}), (\ref{sys_RDL_000000000}),   is  indicated as governing the propensity of system patterning.
 Proceeding, this allows one to infer that  if both outer solutions are unstable, then   instability on both sides of $x=\xi$ is expected. In contrast, if one outer solution is unstable and the other stable, we expect an instability on one side of $x=\xi$. Analogous reasoning suggests stability if the outer solution dynamics either side of $x=\xi$   is stable. Finally, due to this local nature of the result we expect that the spatial frequency of the emerging patterns is also related locally to the Turing conditions and hence with the prospect of a change in spatial frequency of patterning across the domain.


\subsection{Summary and formulation of DDI conditions}

Spectral theory as detailed in the Appendices yields a plausible approach to stability analysis but its practical use seems to be limited as the algebraic complexity even after truncation requires a numerical approach and provides neither information about one-sided patterns nor the effect of spatial heterogeneity on the spatial frequency variation in the resulting pattern. The asymptotics, on the other hand, can be used to estimate conditions for Turing pattern emergence and its classification.

As a summary of all the above partial results suggests that the conjectured conditions are of the form of conditions for Turing's diffusively driven instability evaluated for the system with constant coefficients considered separately on intervals $(0,\xi)$ and $(\xi,L)$. 
Therefore let us denote the following conditions for the  latter interval:
\begin{equation}\label{Tconds_lin}
 	\begin{aligned}
 		T1^R &:= b_{11} +s +b_{22} <0, \\
 		T2^R &:= (b_{11}+s)b_{22} - b_{12}b_{21} >0, \\
 		T3^R &:= (b_{11}+s)d_2 +b_{22}d_1 >0, \\
 		T4^R &:= ((b_{11}+s) d_2 + b_{22}d_1)^2-4d_1d_2((b_{11}+s)b_{22} - b_{12}b_{21}) >0,
 	\end{aligned}
 \end{equation}
 and analogously for the former interval $T1^L$-$T4^L$ (particularly noting that $Ti^L \equiv Ti^R,~i\in\{1,\ldots,4\}$ with $s=0$). These conditions will be compared with results from a large number of numerical solutions to the full model. Further, as we discuss below, we obtain not only a very good approximation for stability conditions in practice but we also observe a strong indication that it provides a tool to distinguish among all the one-sided pattern and both-sided pattern in the case of non-linear kinetics, which is clearly beyond the scope of the spectral analysis in the Appendices.

 In particular assuming $T1^L\land T2^L \land T1^R \land T2^R$ it could be expected that a pattern on the left will emerge only if $T3^L\land T4^L$ holds and the pattern on the right will emerge only if $T3^R\land T4^R$ holds. Therefore we propose and will subsequently numerically verify the following conditions:
\begin{equation}\label{nelin_cond}
	\begin{aligned}
		T1^L&\land T2^L \land T1^R \land T2^R \land \phantom{\neq}(T3^L\land T4^L) \land\phantom{\neq}(T3^R\land T4^R) & &\text{pattern on both sides,}\\
		T1^L&\land T2^L \land T1^R \land T2^R \land \phantom{\neq}(T3^L\land T4^L) \land \neg(T3^R\land T4^R) & &\text{pattern on the left side,}\\
		T1^L&\land T2^L \land T1^R \land T2^R \land \neg(T3^L\land T4^L) \land\phantom{\neq}(T3^R\land T4^R) & &\text{pattern on the right side,}\\
		T1^L&\land T2^L \land T1^R \land T2^R \land \neg(T3^L\land T4^L) \land\neg(T3^R\land T4^R) & &\text{no pattern.}
	\end{aligned}
      \end{equation}
      and we shall use both the spectral approach of the Appendices and numerical solution of the full system to verify these conditions. The former can be used just for the assessment of (in)stability while the latter will be employed to check these conditions for the existence of one-sided pattern. Note that a straightforward extension of these conditions to the general case also be implemented, as discussed in Appendix \ref{lin.general}.

\section{Numerical approach. Verification of estimated DDI conditions}

We shall first focus on the verification of the proposed DDI conditions \eqref{Tconds_lin} in the linear case, where the only plausible prediction and verification is whether a small perturbation exponentially increases or decays in time. Then we proceed to test the more detailed conditions \eqref{nelin_cond} in the nonlinear case where, in addition, we verify the prediction of a one-sided pattern.

\subsection{Linear case}

We have eight conditions \eqref{Tconds_lin} and hence 256 combinations to be analysed. Since we are interested in a phenomenon similar to Turing's self-organisation, we disregard the case when the kinetics themselves induce instability. In classical Turing patterning two of the DDI conditions are equivalent to a requirement of a stable homogeneous steady state in the absence of diffusion \cite{klika2018domain}. 
Therefore, in our case we assume that $T1^L\land T2^L \land T1^R \land T2^R$ holds (corresponding to a stable homogeneous steady state on both parts of $(0,L)$) and we focus on the remaining 16 combinations.

The sets corresponding to each combination are denoted by distinct regions, see Tab.\ \ref{tab_conds}. For all parameters in each region we want to know if a pattern can emerge or not. This crucially includes an assessment of whether one can assign this property of pattern existence to every point in each region, independent of further details. As it was noted in the previous analysis of linear system stability, we will use two tools: (i) calculating the largest real part of eigenvalues of the truncated matrix in the Appendices, Eqn. \eqref{spec_matrix}, using MATLAB and (ii) solving the evolution problem \eqref{sys_RD_lin} using Mathematica.

\begin{table}
	\begin{center}
		\begin{tabular}{|l|c|c|c|c|}
		\hline
		\multicolumn{5}{ | c |}{$T1^L\land T1^R \land T2^L\land T2^R$} \\ \hline
						& $T3^L\land T3^R$ & $\neg T3^L\land T3^R$  & $T3^L\land\neg T3^R$ & $\neg T3^L\land\neg T3^R$  \\ \hline
		$T4^L\land T4^R$			& 	(U,+)		&		(U,+)		&		(U,+)			&		(0,-)				\\ \hline
		$\neg T4^L\land T4^R$		& 	(U,+)		&		(U,+)		&		(0,-)			&		(0,-)				\\ \hline
		$T4^L\land\neg T4^R$		& 	(U,+)		&		(0,-)		&		(U,+)			&		(0,-)				\\ \hline
		$\neg T4^L\land\neg T4^R$	& 	(0,-)		&		(0,-)		&		(0,-)			&		(0,-)				\\ \hline
	\hline
		\multicolumn{5}{ | c |}{$\neg(T1^L\land T1^R \land T2^L\land T2^R)$} \\ \hline
		\multicolumn{5}{ | c |}{(U,+)} \\ \hline
		\end{tabular}
	\end{center}
	\begin{caption}{\label{tab_conds} The table summarising the results for the regions, i.e. the sets of the parameters satisfying combinations of the conditions \eqref{Tconds_lin} in the explored parameter space. All parameters in each region exhibit the same behaviour. We impose the following designation: $U$ and $0$ denote unbounded and zero long-time solutions of the evolution problem, $+$ and $-$ denote signs of the largest real part of eigenvalues of the matrix \eqref{spec_matrix}.}
	\end{caption}
\end{table}

In both approaches we take large sets of parameter values  sampling each region. 
Other parameters are fixed for every numerical experiment if it is not stated otherwise and $L=100$ is chosen to be large enough not to obscure a possible pattern. In particular, from the theory of the classical instability, if the Turing conditions hold then there exists a critical size of the domain such that for every larger domain the system can exhibit a pattern, subject to a weak, discrete, constraint that an integer number of modes is accommodated within the domain. 
The step position $\xi = 30$ is near to, but not exactly at, the domain midpoint, which avoids the precision of symmetry, which, in applications, would not be expected. 

To reduce the seven dimensional parameter space we fix diffusion coefficients with a sufficiently large ratio $d_1 = 1$, $d_2 = 10$, and we also fix $s=0.5$ ($L=100$). 

Before we present the relation between conditions \eqref{Tconds_lin} and pattern formation, we compare the numerical and spectral approaches based on computational results. First, the results from the spectral approach concur with those from solving the evolution problem. In particular, with the possible exception of the very near vicinity of the parameter space boundaries between differing stability behaviours, the largest real part of eigenvalues is negative if and only if the supremum norm of the solution to the evolution problem is smaller than the initial norm. 
Due to the conformity of the results from both methods while being very different conceptually 
the conclusions from either approximation are inferred to be generally accurate.

Second, the character of pattern emergence is indeed the same within each region from Tab \ref{tab_conds}. Particularly, if two regions express an opposite behaviour, the change is located exactly on the border of the regions (with negligible imperfection due to numerical imprecision). Hence, these observations entail a justification of the chosen conditions \eqref{Tconds_lin}.

\begin{figure}
	\centering
	\includegraphics[width=0.45\textwidth]{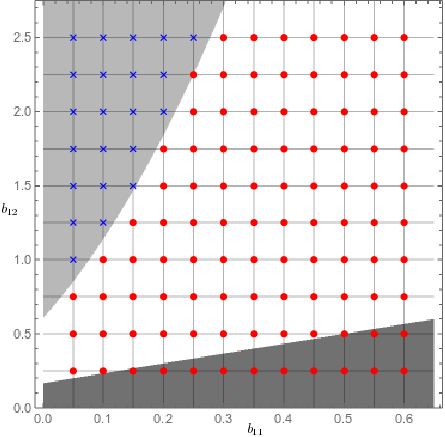}
	\includegraphics[width=0.45\textwidth]{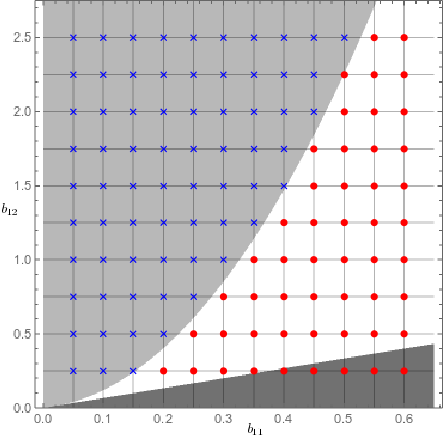}
	\caption{\label{fig_linCond} An illustration of the match between the identified Turing instability conditions for affine kinetics, Eqn. \eqref{lin_cond}, and the results from the evolution of system \eqref{sys_RD_lin} with $b_{10}=1$, $b_{20}=3$, $b_{21}=-3$, $b_{22}=-2$, $L=100$, $\xi=30$, $d_1=1$, $d_2=100$ and: A) $s=0.25$, B) $s=-0.25$. The remaining two parameters, $b_{11},~b_{12}$, are considered as parameters for exploring stability properties and are on the $x$ and $y$ axes. In the background the conditions \eqref{lin_cond} are plotted in the grayscale, 
          which, in increasing grayscale intensity, highlight regions with an unbounded solution indicating the existence of pattern (white), decaying solution indicating no pattern and the region where $T1^L\land T2^L \land T1^R \land T2^R$ does not hold (the darkest grey). The spots denote the resulting pattern type based on numerical solution to the evolution problem: an unbounded solution indicating a pattern (a red disk) and decaying solutions corresponding to no pattern (a blue cross).}
      \end{figure}

The results are outlined in Tab. \ref{tab_conds} and Fig \ref{fig_linCond} which  can be summarised as: a Turing pattern will emerge for large enough $\text{min}(\xi,L-\xi)$ with unstable eigenmodes that satisfy the boundary conditions if and only if
\begin{equation}\label{lin_cond}
	\big(T1^L\land T2^L \land T1^R \land T2^R\big) \land \bigg( \big(T3^L\land T4^L\big) \vee \big(T3^R\land T4^R\big) \bigg)
\end{equation}
holds.

In short, the hypothetical conditions \eqref{Tconds_lin} have been evidenced as well-characterising whether the system exhibits a pattern or not. Even though we have implemented a detailed numerical verification using two different approaches -- computing a spectrum of the truncated matrix \eqref{spec_matrix} and solving the evolution problem \eqref{sys_RD_lin} -- and these conditions seem to have an intuitive explanation, we are not able to validate them nor relate them using rigorous analytical approach, as we indicated above. However, one should bear in mind that the analysis outlined in the Appendices via a spectral analysis is lacking rigour ``only'' in the limit $N\rightarrow \infty$ where $N$ is the number of terms in the spectral expansion, while the continuum approximation motivates a finite $N$ cut-off. 


\subsection{Nonlinear case}

Additionally, from \eqref{nelin_cond} we have the prospect of (i) a tool to distinguish a one-sided pattern from a both-sided pattern and (ii) an indicative criterion for self-organisation even for non-linear kinetics, that is how to find conditions determining pattern emergence, and (iii) how to detect which type of the pattern it should be. We proceed to test this, considering Schnakenberg kinetics
\begin{equation}\label{kin_Sch}
	\begin{aligned}
		f(u,v) &= a-u +u^2v, & g(u,v) &= b-u^2v,
	\end{aligned}
\end{equation}
and Gierer-Meinhardt kinetics
\begin{equation}\label{kin_GM}
	\begin{aligned}
		f(u,v) &= a-bu+\frac{u^2}{v}, & g(u,v) &= u^2-v,
	\end{aligned}
\end{equation}
with $a$,$b$ positive constants as two exemplars for reaction kinetics in Turing models.

Numerical experiments have been implemented using Wolfram Mathematica as in the linear case. 
The terminal time is $\tau=10^3$. This choice was sufficient to distinguish the non-existence of pattern from its  presence, where in the latter case the convergence of a norm was clearly observed suggesting a convergence of the long-time solutions to stationary patterns. The initial condition was set to be  small random noise around the stationary solution $(\bar{u}(x), \bar{v}(x))$.
For both choices of kinetics we take $L=400$, $\xi=120$, $d_1=1$, $d_2=100$; this parameter selection follows the reasoning from the linear case. Large sets of the remaining parameters $a,b,s$ are considered to capture the rich behaviour sufficiently to illustrate the legitimacy of the instability conditions  \eqref{nelin_cond}.

\begin{figure}
	\centering
	\includegraphics[width=0.45\textwidth]{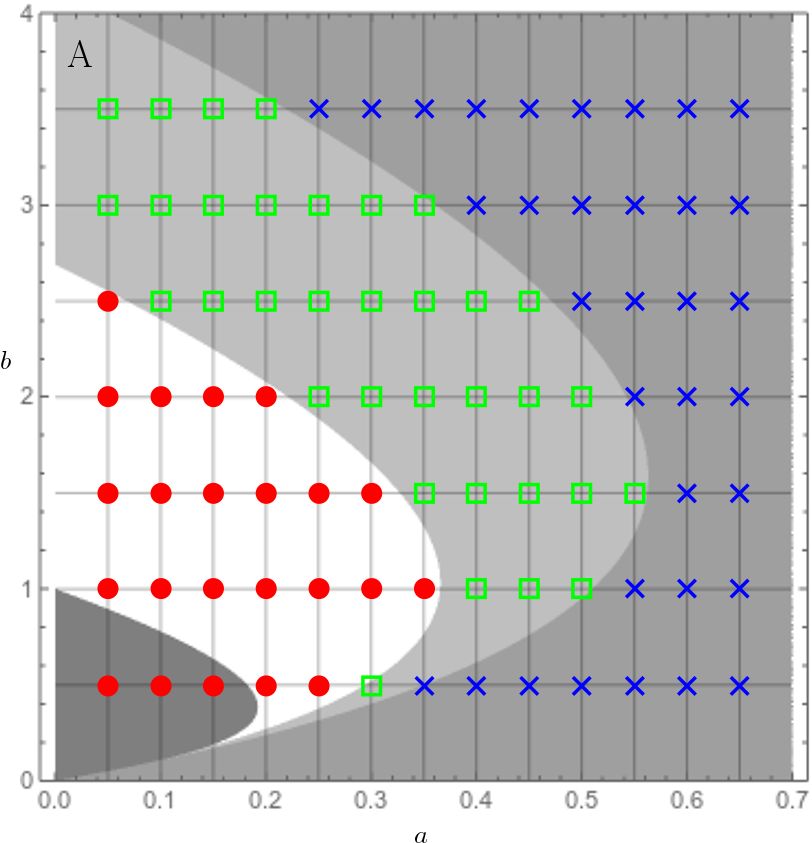}
	\includegraphics[width=0.45\textwidth]{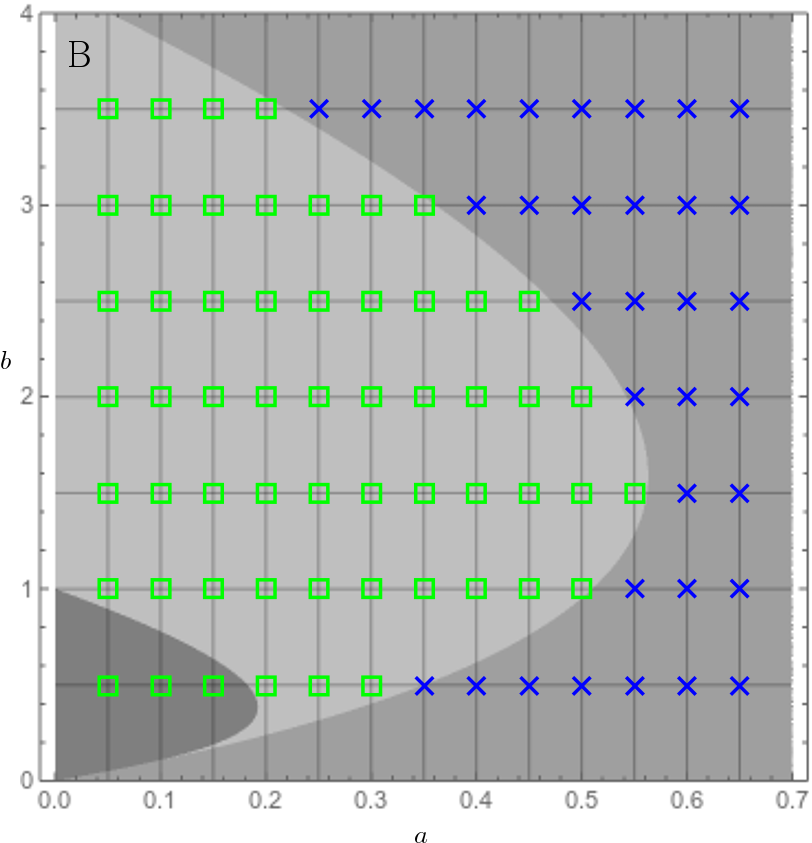}
	\includegraphics[width=0.45\textwidth]{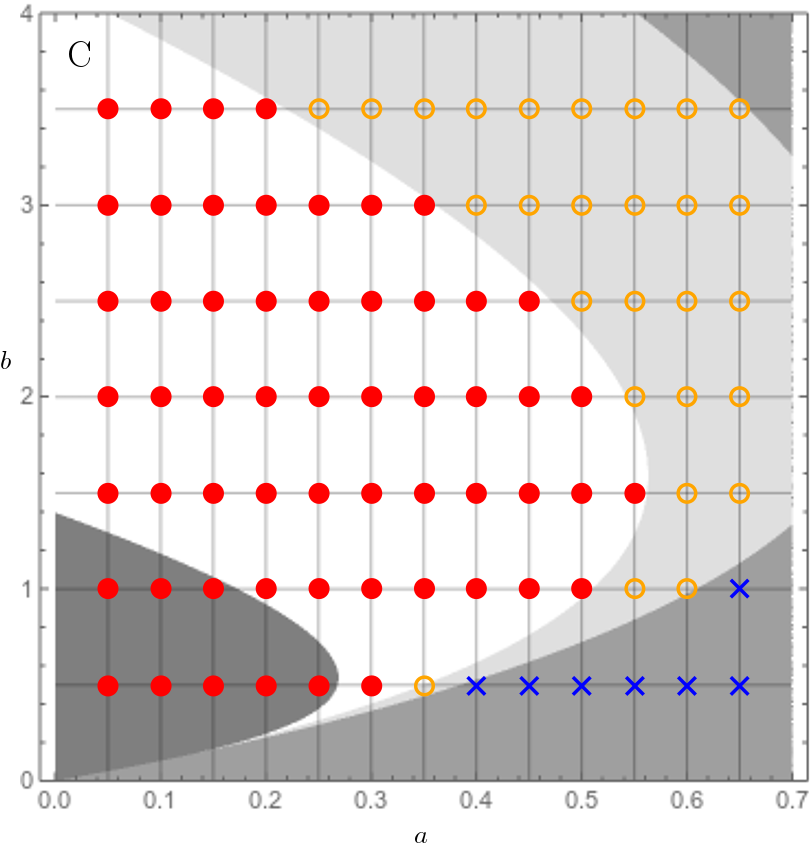}
	\includegraphics[width=0.45\textwidth]{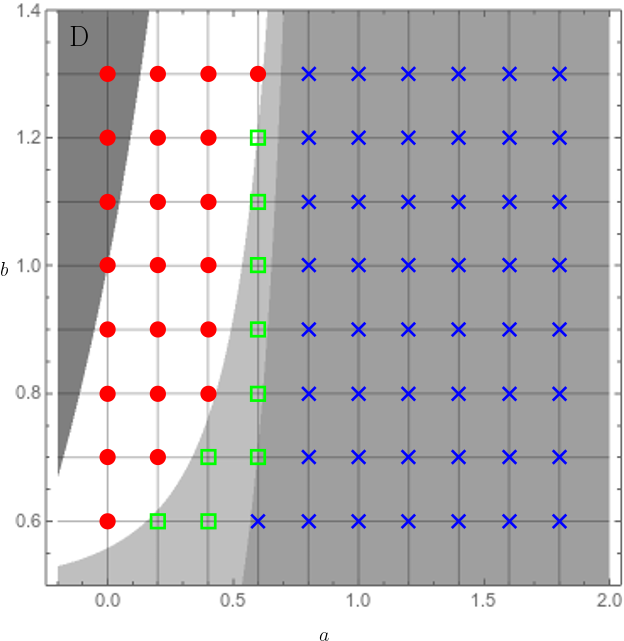}
	\caption{\label{fig_nelinCond} An illustration of the match between the identified Turing instability conditions, Eqn. \eqref{nelin_cond}, and the results from the evolution of system \eqref{sys_RD} with Schnakenberg kinetics: A) $s=0.25$, B) $s=0.75$, C) $s=-0.25$; and Gierer-Meinhardt kinetics: D) $s=0.5$. In the background the conditions \eqref{nelin_cond} are plotted in the grayscale, which, in increasing grayscale intensity, highlight regions with a both-sided pattern (white), a right-sided pattern, left-sided pattern, no pattern and the region where $T1^L\land T2^L \land T1^R \land T2^R$ does not hold (the darkest grey). The spots denote the resulting pattern type based on numerical solution to the evolution problem: a both-sided pattern (a red disk), a left-sided pattern (a green square), a right-sided pattern (an orange circle) and no pattern (a blue cross).}
\end{figure}

In particular, the types of pattern resulting from simulations agree well with the predictions given by conditions \eqref{nelin_cond} as depicted in Fig. \ref{fig_nelinCond}. The degree of correspondence seems to be very high at least in the tested scenarios (kinetics and parameters selection) giving merit to the approach and the resulting conditions, despite the absence of rigour.

\section{Conclusions}\label{sec_disc}
\label{Concl_Summary}
In this paper we considered a reaction diffusion system with a  spatial dependence via a linear kinetic term with a coefficient in the form of a spatial step function and we analysed the resulting impact on conditions for pattern formation. First we defined a pattern as a steady solution with an inhomogeneity persevering throughout a  large enough domain. Using an analytical-numerical approach we examined a case of affine kinetics and deduced conditions for pattern emergence in a very simple form, Eqn. \eqref{Tconds_lin}. For the case of non-linear kinetics we took conditions inherited from a suitable linearisation, generalising the previous conditions to those stated in \eqref{nelin_cond}, with a verification for two choices of kinetics and a range of parameter values. Further note that conditions \eqref{nelin_cond} and their agreement with numerics also match the intuition of at least some experimentalists in the field, e.g. \cite{miguez2006effect}, and suggests further analytical progress may be feasible at least for step function behaviours in the kinetics.


If we compare our results with the previously studied system with an additive spatial step function independent of the morphogen concentrations \cite{page2003}, the patterns in our case have not only different amplitudes on the two sides of the step but also different frequencies. 
This highlights that patterning with sharp changes in spatial frequency may be a signature of kinetic heterogeneity characterised by rapid transitions in kinetics.

\label{Concl_DependenceOnOtherCoefficients}
One interesting point is whether there is any restriction on the size of the step $s$. Although the final conditions \eqref{nelin_cond} are well defined for any value of $s$, a restriction arises from our definition of a pattern in Section \ref{sec_intro}.
We need that the eventual pattern should be more significant in comparison to the inhomogeneity localised around $\xi$, which is expected to be true for sufficiently small $s$, by appeal to continuity with respect to parameters. For larger $s$ the localised inhomogeneity will also be larger, which can be easily seen even in the linear case due to the larger gap between $\bar{u}(x)$ (resp. $\bar{v}(x)$) at the point $\xi$ and thus has every potential to invalidate our findings for sufficiently large $s$.

The conditions \eqref{nelin_cond} are necessary diffusion driven instability (DDI) conditions and depend directly on the diffusion coefficients, kinetics and the size of the step. As well as in the classical Turing system, if DDI conditions hold, a large enough domain is necessary for pattern to emerge; thus the intervals $(0,\xi)$ and $(\xi,L)$ are also required to be sufficiently large.
Moreover, sufficiently large intervals are necessary for a pattern to be correctly identifiable, as seen in the comparison of Fig. \ref{fig_res1}B and Fig. \ref{fig_res2}B 
where both systems are predicted to exhibit patterning on the left of the step, but only the latter does, since the subdomain $[0,\xi)$ is smaller than the emergent pattern period in Fig. \ref{fig_res1}B.
Finally, note that differences in spatial frequencies are observed to be independent of both interval lengths and hence independent of both $\xi$ and $L$. However, as the boundary layer analysis suggests, the spatial frequency can be different in the two parts of the domain as it is evaluated independently on the two subintervals.

\label{Concl_generalisation}
This article is concerned with a special case of heterogeneity in the kinetics though this can be easily generalised to a certain extent. Firstly, it is easy to see that the particular choice of a spatially dependent linear term in the kinetics is not important for the analysis even in the case of non-linear kinetics. Further, the same approach as well as the results, will be valid for a step function $h(x)$ with finitely many steps. A limitation arises, however, due to the note from the previous paragraphs -- the sizes of the steps should be small enough in comparison to the surrounding intervals so that the localised inhomogeneity due to the step does not exceed the emerging pattern in magnitude/amplitude. Further, the discreteness of eigenmodes results in a lower bound on the size of the supporting intervals of each step.

The question of a generalisation to a spatial dependency is expected for slowly varying function $h(x)$. In such a case we can take an approximation of $h(x)$ using a simple function and take the advantage of conclusions of this approximative system. In the case of a more general dependency it may be difficult to show similar conclusions. Actually, as a recent study shows \cite{Krause2018PRE} even shallow gradients coupled to non-linear kinetics may lead to an unexpected and complex behaviour but finding a clear cut distinction between these cases is beyond the scope of this paper. Furthermore, how and where the present intuitive analysis fails also remains to be explored as do higher dimensional domains and curved geometries, which may allow ready generalisation.

We could also consider a higher-dimensional space. The presented approach is easily extendable to domains in the form of higher-dimensional rectangles. For example, one might readily find conditions yielding the emergence of a pattern with spots on one part and stripes on the other part of a higher dimensional domain.

Finally, let us finish with a summary of the identified hypothetical DDI conditions: (in)stability in the Turing model analysed here appears to be a local property and can be analysed as such, with the local assessment of whether parameters are within Turing space providing a strong indication for an unstable eigenmode excitation, at least on a sufficiently large domain with concomitant spatial frequency heterogeneity.

\section*{Acknowledgement}
M.K. has been supported by the Fog Research Institute under contract no.~FRI-454.
V.K. is grateful for support from the International Mobility of Researchers--MSCA-IF in Czech Technical University grant CZ.02.2.69/0.0/0.0/17\_050/0008025 funded by The Ministry of Education, Youth and Sports (MEYS) of the Czech Republic, European Regional Development Fund-Project ``Center for Advanced Applied Science'' (No. CZ.02.1.01/0.0/0.0/16 019/0000778) and the Mathematical Institute at the University of Oxford.


\appendix

\section{Further details of stability analysis}

In the main text, Section \ref{sect_linStab}, we followed a rather standard approach to linear stability analysis and simply adopt it for a system with a jump in reaction kinetics. Here, for the sake of completeness, we pursue other possible approaches and, as we shall show, all entail the same qualitative behaviour. Namely, we expand the jump in a generalised function sense, see Appendix \ref{App.delta}, and linearise just the kinetics instead of linearisation of the whole problem, see Appendix \ref{App.LinearisationOfKinetics}.

These approaches lead to problem of the type Eqn. \eqref{Eq.12} which can be further analysed using spectral methods, i.e. using expansions in eigenfunctions of negative Laplacian, see Appendix \ref{sec_lin} for details. However, the system does not decouple into a straightforward collection of coupled ordinary differential equations for amplitudes of the eigenfunctions but rather a truncation has to be employed. Finally, an extension to a general linear case with jumps in every term is straightforward as noted in Appendix \ref{lin.general}. Note that all computational results shown in this work were obtained via two approaches which are described in Appendix \ref{app.compu}.


\subsection{Alternative I. Linearisation of kinetics only} \label{App.LinearisationOfKinetics}

An alternative approach is to linearise only the kinetics around the piecewise constant naive steady state. 

Assume the existence of a solution $(\bar{u}^L, \bar{v}^L)$ and $(\bar{u}^R, \bar{v}^R)$ satisfying
\begin{equation*}
	\begin{aligned}
		f(\bar{u}^L, \bar{v}^L) &=0 = g(\bar{u}^L, \bar{v}^L) \\
		f(\bar{u}^R, \bar{v}^R) +s\bar{u}^R &=0 = g(\bar{u}^R, \bar{v}^R)
	\end{aligned}
\end{equation*}
and take approximations of reaction kinetics as Taylor expansions evaluated separately on both intervals ($\tilde{u} = u-\bar{u}$):
\begin{equation*}
	\begin{aligned}
		L&: & f(u,v) &= f(\bar{u}^L, \bar{v}^L) + J_{11}^L\tilde{u} + J_{12}^L\tilde{v} + \dots & 
					g(u,v) &= g(\bar{u}^L, \bar{v}^L) + J_{21}^L\tilde{u} + J_{22}^L\tilde{v} + \dots \\
		R&: & f(u,v)+su &= f(\bar{u}^R, \bar{v}^R) + s\bar{u}^R+ J_{11}^R\tilde{u} + J_{12}^R\tilde{v} + \dots & 
					g(u,v) &= g(\bar{u}^R, \bar{v}^R) + J_{21}^R\tilde{u} + J_{22}^R\tilde{v} + \dots
	\end{aligned},
\end{equation*}
where $J_{ij}^L$ (resp. $J_{ij}^R$) denote the elements of Jacobian matrix of the map $(f,g)$ at $(\bar{u}^L, \bar{v}^L)$ (resp. $(f+su,g)$ at $(\bar{u}^R, \bar{v}^R)$). Using the following notation
\begin{equation*}
	J_{ij}(x)=\begin{cases}
		J_{ij}^L & x \in (0,\xi) \\
		J_{ij}^R & x \in (\xi,L).
		\end{cases},\quad
	\bar{u}(x) = \begin{cases}
		\bar{u}^L & x \in (0,\xi) \\
		\bar{u}^R & x \in (\xi,L)
		\end{cases}, \quad
	\bar{v}(x) = \begin{cases}
		\bar{v}^L & x \in (0,\xi) \\
		\bar{v}^R & x \in (\xi,L)
		\end{cases}
\end{equation*}
we can write down an affine system describing evolution around $(\bar{u}(x), \bar{v}(x))$ while approximating the original system with non-linear kinetics as
\begin{equation}\label{sys_RDL}
	\begin{aligned}
		\partial_t u &= d_1 \partial_{xx}u + J_{11}(x)u + J_{12}(x)v +c_1(x)\\
		\partial_t v &= d_2 \partial_{xx}v + J_{21}(x)u + J_{22}(x)v +c_2(x)
	\end{aligned}\quad\text{on $(0,L)$,}
\end{equation}
with the step functions
\begin{equation*}
	c_1(x)=-J_{11}(x)\bar{u}(x) - J_{12}(x)\bar{v}(x), \quad c_2(x)=-J_{21}(x)\bar{u}(x) - J_{22}(x)\bar{v}(x).
\end{equation*}

The analysis of such a system in the current framework is discussed in Appendix \ref{lin.general}.


\subsection{Alternative II. Linearisation about a piecewise constant steady state} \label{App.delta}
The second option could be to linearise around step functions $(\bar{u}, \bar{v})$. If we try to proceed, we will obtain a linear system, but with terms containing a derivative of the Dirac delta function, which results from the non-trivial step in $(\bar{u}, \bar{v})$. Therefore, we are not able to obtain the linearised system following the standard approach above.

However, it is instructive to proceed further as an expansion (in generalised functions) of the delta function in the eigenfunctions $\{y_k\}$ is available and hence we can rewrite the linearised system yet again in terms of a system of equations for particular modes.

Linearisation of the system \eqref{sys_RD} around step functions $(\bar{u}(x), \bar{v}(x))$ (defined in \eqref{def_PSS}) is well defined in the distributional sense and is of the form
\begin{equation}\label{app_sys_RD_lin}
	\begin{aligned}
		\partial_t \tilde{u} &= d_u \partial_{xx}\tilde{u} + b_{11}\tilde{u} +b_{12}\tilde{v} + s_u d_u \delta'(x-\xi)\\
		\partial_t \tilde{v} &= d_v \partial_{xx}\tilde{v} + b_{21}\tilde{u} +b_{22}\tilde{v} + s_v d_v \delta'(x-\xi)
	\end{aligned}
	\quad \text{in $(0,L)$,}
\end{equation}
where $(s_u, s_v)$ denotes the sizes of the step of $(\bar{u}(x), \bar{v}(x))$ at $\xi$,  $\delta(x)$ denotes Dirac delta function and  $u(x) = \tilde{u}(x) - \bar{u}(x)$, $v(x) = \tilde{v}(x)-\bar{v}(x)$.

Since Neumann boundary conditions are considered, we expand $(\tilde{u}, \tilde{v})$ using orthonormal basis $\{y_n\}_{n\in\{0,1,..\}} = \left\{\frac{1}{L}, \frac{2}{L} \cos{\left(\frac{n\pi}{L}x\right)}\right\}_{n=1}^\infty$ as the series
\begin{equation}
	\tilde{u} = \sum_{n=0}^\infty A_n y_n, \quad 
    \tilde{v} = \sum_{n=0}^\infty B_n y_n
\end{equation}
and rewrite system \eqref{app_sys_RD_lin} in the form of a system of equations for each eigenmode. The Dirac delta function can be expanded in terms of any eigenfunctions of the Laplacian on any interval. Hence we use the following expansion of Dirac delta function on $(0,L)$
\begin{equation*}
  \delta(x-\xi) =\frac{2}{L} \sum_{n=1}^{\infty} \sin\left(\frac{n\pi \xi}{L}\right) \sin\left(\frac{n \pi x}{L}\right).
\end{equation*}
Therefore the linearised problem has the following eigenmode expansion
\begin{equation} \label{25half}
  \sum_{n=1}^\infty y_n \left\{\begin{pmatrix}\dot{A}_n \\ \dot{B}_n\end{pmatrix} + \left[\begin{pmatrix} d_u & 0\\ 0 & d_v\end{pmatrix} \kappa_n - \mathbf{J}(x) \right]\begin{pmatrix}A_n \\ B_n\end{pmatrix}   - \underbrace{\frac{n\pi}{L} \sin\left(\frac{n\pi \xi}{L}\right) \begin{pmatrix} d_u s_u & 0 \\ 0 & d_v s_v\end{pmatrix}\begin{pmatrix}A_n \\ B_n\end{pmatrix}}_{\mbox{forcing}}\right\} = 0,
\end{equation}
where $\kappa_n = \left(\frac{n \pi}{L}\right)^2$ and the matrix of linearised kinetics $\mathbf{J}(x)$ is evaluated at the piece-wise constant function $(\bar{u}(x), \bar{v}(x))$. As the $\delta'(x-\xi)$ contribution translates only into a (constant) forcing, it does not affect the (in)stability result. Thence the generalised function approach yields exactly the same problem as previously derived in Eqn. \eqref{Eq.12}.


\subsection{Detailed analysis of the dispersion relation}\label{sec_lin}

The difference from the standard Turing system analysis for a homogeneous system emerges from the spatial dependence of $\BJ(x)$ preventing the decoupling of individual eigenmodes and hence preventing straightforward solution. However, we can take the advantage of the fact that $\BJ(x)$ contains only constants and a step function, all satisfying Neumann boundary conditions and hence within the span of the eigenfunctions $\{y_k\}$
\begin{equation}
	\BJ(x) = \sum_{k=0}^\infty \underbrace{\begin{pmatrix} J_{11}^{(k)} & J_{12}^{(k)} \\ J_{21}^{(k)} & J_{22}^{(k)}\end{pmatrix}}_{\BJ^{(k)}}y_k(x).
\end{equation}
The system can be rewritten as
\begin{equation} \label{27}
	\sum_{n=0}^\infty \left[\begin{pmatrix} \partial_t A_n \\ \partial_t B_n\end{pmatrix} y_n(x)+ \BD\begin{pmatrix}A_n \\ B_n\end{pmatrix} \kappa_n y_n(x) - \sum_{k=0}^\infty \underbrace{\begin{pmatrix} J_{11}^{(k)} & J_{12}^{(k)} \\ J_{21}^{(k)} & J_{22}^{(k)}\end{pmatrix}\begin{pmatrix}A_n \\ B_n\end{pmatrix}}_{=: \BC_{k,n}} y_k(x) y_n(x) \right]= 0.
\end{equation}
The eigenfuctions of the negative Laplacian on a one-dimensional interval are of the well-known form $y_n(x)=\cos{(n\pi x/L)}$ and hence we have
\begin{equation}
	\begin{aligned}
		y_k(x) y_n(x) &= \frac{1}{2}\left(\cos{\frac{(n+k)\pi x}{L}} + \cos{\frac{(n-k)\pi x}{L}}\right) = \frac{1}{2}\left(\cos{\frac{(n+k)\pi x}{L}} + \cos{\frac{|n-k|\pi x}{L}}\right) = \\
		&= \frac{y_{n+k}(x) + y_{|n-k|}(x)}{2},
	\end{aligned}
\end{equation}
which are again functions from the orthogonal basis. To obtain the dispersion relation we need to reorder the second sum to be able to factor out the function $y_n(x)$ and then invoke orthogonality of the orthogonal basis to transform the problem into an infinite system of ordinary differential equations. Denoting the coefficients in the internal sum by $\BC_{k,n} \in\mathbb{R}^2$ we obtain the following form of the system:
\begin{equation}
	\begin{aligned}
		\sum_{m=0}^\infty \begin{pmatrix} \partial_t A_m \\ \partial_t B_m\end{pmatrix} &y_m(x)+ \BD\begin{pmatrix}A_m \\ B_m\end{pmatrix} \kappa_m y_m(x) =\\
		&= \frac{1}{2}\sum_{m=0}^\infty \left( \sum_{n=0}^m \BC_{m-n,n}+\sum_{n=m}^\infty \BC_{n-m,n}\right) y_m(x) + \frac{1}{2}\sum_{m=1}^\infty\sum_{n=0}^\infty \BC_{n+m,n} y_m(x).
	\end{aligned}
\end{equation}

The coupled evolution equations for the eigenmodes are then of the form:
\begin{equation}\label{RDSlin_eqModes}
	\begin{aligned}
		0&= \begin{pmatrix} \partial_t A_m \\ \partial_t B_m\end{pmatrix} + \BD \begin{pmatrix}A_m\\B_m\end{pmatrix} \kappa_m - \frac{1}{2} \sum_{n=0}^\infty \BC_{|m-n|,n} -\frac{1}{2}\sum_{n=0}^\infty \BC_{m+n,n} - \frac{1}{2}\BC_{0,m}, \quad \text{for } m\geq 1,\\
		0&=\begin{pmatrix} \partial_t A_0 \\ \partial_t B_0\end{pmatrix} + \BD \begin{pmatrix}A_0\\B_0\end{pmatrix} \kappa_0 - \frac{1}{2}\sum_{n=0}^\infty \BC_{n,n}- \frac{1}{2}\BC_{0,0}.
	\end{aligned}
\end{equation}

In our case, the elements of matrices $\BJ^{(k)}$ can be computed as:
\begin{equation}\label{lin_J_fourcoef}
	\BJ^{(k)} = \begin{pmatrix}J_{11}^{(k)} & J_{12}^{(k)}\\J_{21}^{(k)} & J_{22}^{(k)}\end{pmatrix} =
		\begin{cases}
			\begin{pmatrix}Z_k & 0\\0 & 0\end{pmatrix} & k\ge 1 \\
			\begin{pmatrix}Z_0 & b_{21} \\ b_{21} & b_{22}\end{pmatrix} & k=0,
		\end{cases}
\end{equation}
where the spatial heterogeneity is represented by a step function and where 
\begin{equation}
	\begin{aligned}\label{lin_J_fourcoef2}
        Z_k &= \frac{\langle h(x), y_k(x) \rangle}{\|y_k(x)\|^2} = \frac{2}{L}\int_\xi^L s\cos{\frac{k\pi x}{L}} \d x = -\frac{2s}{k\pi} \sin{\frac{k\pi \xi}{L}},\\
		Z_0 &= \frac{\langle h(x)+b_{11}, y_0(x) \rangle}{\|y_0(x)\|^2} = b_{11} + \frac{1}{L}\langle h(x),1\rangle = b_{11} + \frac{s(L-\xi)}{L}.
	\end{aligned}
\end{equation}

While in the case of spatial homogeneity, the spectrum and dispersion relation for the system rate of growth in terms of wavenumber is given by the solvability condition for the eigenmodes, the analogous information is not analytically accessible in this framework for spatially heterogeneous functions.


Nevertheless, the system \eqref{RDSlin_eqModes} is linear and hence the solution can be written in terms of an exponential of a linear operator. Since we are not able to calculate the spectrum of the infinite matrix
\begin{equation}\label{spec_matrix}
	\begin{pmatrix}
    Z_0-d_1\kappa_0 & b_{12}    & \frac{Z_1}{2} & 0                      & \frac{Z_2}{2} & 0                      & \dots \\
    b_{21} & b_{22}-d_2\kappa_0 & 0     & 0                              & 0     & 0                              & \dots \\
    Z_1   & 0                   & Z_0+\frac{Z_2}{2}-d_1\kappa_1 & b_{12} & \frac{Z_1+Z_3}{2} & 0                  & \dots \\
    0     & 0                   & b_{21} & b_{22}-d_2\kappa_1            & 0     & 0                              & \dots \\
    Z_2   & 0                   & \frac{Z_1+Z_3}{2} & 0                  & Z_0+\frac{Z_4}{2}-d_1\kappa_2 & b_{12} & \dots \\
    0     & 0                   & 0     & 0                              & b_{21} & b_{22}-d_2\kappa_2            & \dots \\
    Z_3   & 0                   & \frac{Z_2+Z_4}{2} & 0                  & \frac{Z_1+Z_5}{2} & 0                  & \dots \\
    0     & 0                   & 0     & 0                              & 0     & 0                              & \dots \\
    \vdots & \vdots & \vdots & \vdots & \vdots & \vdots & \ddots
	\end{pmatrix}
\end{equation}
we will use MATLAB to estimate it by calculating spectrum of its truncated principal submatrix $\mathbf{M}_n\in\BBC^{2n,2n}$. First let us explore some properties of the infinite matrix. It can be rewritten via the following synoptic sum of two matrices
\begin{equation}\label{spec_matrix_2}
	\begin{pmatrix}
		0			& \frac{Z_1}{2}		& \frac{Z_2}{2} 	& \frac{Z_3}{2} 		& \frac{Z_4}{2} 		& \dots \\
		Z_1			& \frac{Z_2}{2}		& \frac{Z_1+Z_3}{2}	& \frac{Z_2+Z_4}{2} 	& \frac{Z_3+Z_5}{2} 	& \dots \\
		Z_2			& \frac{Z_1+Z_3}{2} & \frac{Z_4}{2}		& \frac{Z_1+Z_5}{2} 	& \frac{Z_2+Z_6}{2} 	& \dots \\
		Z_3			& \frac{Z_2+Z_4}{2} & \frac{Z_1+Z_5}{2}	& \frac{Z_6}{2} 		& \frac{Z_1+Z_7}{2} 	& \dots \\
		Z_4			& \frac{Z_3+Z_5}{2} & \frac{Z_2+Z_6}{2} & \frac{Z_1+Z_7}{2}		& \frac{Z_8}{2} 		& \dots \\
		\vdots & \vdots & \vdots & \vdots & \vdots & \ddots
	\end{pmatrix}\otimes \begin{pmatrix} 1 & 0 \\ 0 & 0 \end{pmatrix} + \bigoplus_{i=0}^\infty
	\begin{pmatrix}
		Z_0-d_1\kappa_i	& b_{12} \\
		b_{21}	& b_{22}-d_2\kappa_i
	\end{pmatrix},
      \end{equation}
      where $\otimes$ stands for the Kronecker product and $\oplus$ denotes a direct sum.

Since $Z_i$ denotes a Fourier coefficient and its norm vanishes for $i\rightarrow \infty$, the first matrix is bounded, compact and has a high degree of symmetry and therefore can be intuitively understood as a small perturbation of the second matrix which is unbounded since $\kappa_i$ grows to infinity as $i\rightarrow \infty$. 
We will not be able to show that the spectrum of the infinite matrix is a limit of the spectrum of the truncated matrices but we shall show that the stability of the truncated linear system is determined by spectrum of a matrix of a relatively small dimension. Further we may justify truncating the matrix due to the continuum approximation, which is behind the formulation of the model itself.

\subsubsection{Spectra of truncated matrices} \label{AppA31}
We shall show that with $\sigma(\mathbf{M}_{N+1})$ denoting the spectrum of the truncated matrix $\mathbf{M}_{N+1}$
\begin{equation*}
	\begin{pmatrix}
    Z_0-d_1\kappa_0 & b_{12}    & \frac{Z_1}{2} & 0                      & \dots & \frac{Z_N}{2} & 0                      \\
    b_{21} & b_{22}-d_2\kappa_0 & 0     & 0                              & \dots & 0     & 0                              \\
    Z_1   & 0                   & Z_0+\frac{Z_2}{2}-d_1\kappa_1 & b_{12} & \dots & \frac{Z_{N-1}+Z_{N+1}}{2} & 0          \\
    0     & 0                   & b_{21} & b_{22}-d_2\kappa_1            & \dots & 0     & 0                              \\
    Z_2   & 0                   & \frac{Z_1+Z_3}{2} & 0                  & \dots & \frac{Z_{N-2}+Z_{N+2}}{2} & 0          \\
    0     & 0                   & 0     & 0                              & \dots & 0     & 0                              \\
    \vdots & \vdots & \vdots & \vdots & \ddots & \vdots & \vdots                                                          \\
    Z_N   & 0                   & \frac{Z_{N-1}+Z_{N+1}}{2} & 0          & \dots & Z_0+\frac{Z_{2N}}{2}-d_1\kappa_{N} & b_{12}\\
    0     & 0                   & 0     & 0                              & \dots & b_{21} & b_{22}-d_2\kappa_{N}            
	\end{pmatrix}
      \end{equation*}
it holds that $ \sigma(\mathbf{M}_{N+1}) \approx \sigma(\mathbf{M}_N)\cup \{-d_1 \kappa_N+O(1),-d_2\kappa_N+O(1)\}$ as $N\rightarrow \infty$.

In particular, we shall show that with $N$ large enough, two eigenvalues are of the order $\kappa_N$ (which grows to infinity as $N\rightarrow\infty$). With $\lambda = \mu \kappa_N$, $\mu=O(1)$ as $N\rightarrow \infty$ we have that
\begin{multline*}
  \det(\mathbf{M}_{N+1}-\lambda\mathbf{I})=\kappa_N^2 \Bigg\{\left[\det(\mathbf{M}_N-\lambda\mathbf{I}) \left(\frac{1}{\kappa_N} (Z_0+Z_{2N}/2) - d_1 -\mu\right)+O(\kappa_N)^{2N-1} \right] \\
    \times\left(\frac{b_{22}}{\kappa_N}-d_2-\mu\right)+\frac{b_{12}}{\kappa_N} O(\kappa_N)^{2N}\Bigg\}=\\
     =\det(\mathbf{M}_N-\lambda\mathbf{I}) ( b_{22} - \kappa_N(d_2+\mu))\left(Z_0+\frac{Z_{2N}}{2} - \kappa_N (d_1+\mu)\right) + O(\kappa_N)^{2N+1},
\end{multline*}
where we note that the $\det(\mathbf{M}_N)$ is a polynomial of $2N$-th order. Therefore two eigenvalues are indeed of the order $\kappa_N$, in particular $\lambda_1= - d_1 \kappa_N+O(1)$ and $\lambda = -d_2 \kappa_N+O(1)$, while the remaining $2N$ eigenvalues are (in the leading order) the eigenvalues of $\mathbf{M}_N$. Hence the information about the stability associated with an arbitrarily large truncated matrix can be deduced from a smaller matrix $\mathbf{M}_N$ (in practice the choice of $N=50$ seems to be good enough). In addition, we anticipate that this characteristic of the spectrum will translate even into the arbitrarily large $N$ case and which seems to be confirmed by the numerical calculations in the main text.

\subsection{Spatial dependence in other kinetics coefficients} \label{lin.general}
We now consider the linear system $\eqref{sys_RD_lin}$ but with spatial dependence in every kinetic coefficient. This dependence is in the form of a step function with various stepsizes but located at the same point $\xi$. Thus consider
\begin{equation}\label{sys_RD_linFull}
	\begin{aligned}
		\partial_t u &= d_1 \partial_{xx}u + b_{10}(x)+b_{11}(x)u +b_{12}(x)v,\quad \mbox{ in } (0,L)\\
		\partial_t v &= d_2 \partial_{xx}v + b_{20}(x)+b_{21}(x)u +b_{22}(x)v,\quad \mbox{ in } (0,L)
	\end{aligned}
\end{equation}
with
\begin{equation*}
	b_{ij}(x)=\begin{cases}
		b_{ij} & x \in [0,\xi)\\
		b_{ij}+s_{ij} & x \in [\xi,L]
	\end{cases} \qquad i \in \{1,2\}, j \in \{0,1,2\}
\end{equation*}
and we assert that the procedure and conclusion are the same as in the above. Indeed, let us briefly repeat the procedure.
The system \eqref{sys_RD_linFull} has constant coefficients if considered separately on intervals $(0,\xi)$ and $(\xi,L)$, therefore the steady state can be expressed at least in principle. The stability analysis proceeds in the same way, only the Fourier coefficients $\BJ^{(k)}$, eq. \eqref{lin_J_fourcoef}, are of the form
\begin{equation*}
	\begin{aligned}
		J_{ij}^{(k)} &= \frac{\langle h(x), y_k(x) \rangle}{\|y_k(x)\|^2} = \frac{2}{L}\int_\xi^L s_{ij}\cos{\frac{k\pi x}{L}} \d x = -\frac{2s_{ij}}{k\pi} \sin{\frac{k\pi \xi}{L}}\\
		J_{ij}^{(0)} &= \frac{\langle b_{ij}(x), y_0(x) \rangle}{\|y_0(x)\|^2} = b_{ij} + \frac{s_{ij}(L-\xi)}{L}
	\end{aligned}
\end{equation*}
and the stability matrix 
\begin{equation}\label{spec_matrix_full}
	\begin{pmatrix}
		0			& \frac{\BJ^{(1)}}{2}			& \frac{\BJ^{(2)}}{2} 			& \frac{\BJ^{(3)}}{2} 				& \frac{\BJ^{(4)}}{2} 			& \dots \\
		\BJ^{(1)}	& \frac{\BJ^{(2)}}{2}			& \frac{\BJ^{(1)}+\BJ^{(3)}}{2}	& \frac{\BJ^{(2)}+\BJ^{(4)}}{2} 	& \frac{\BJ^{(3)}+\BJ^{(5)}}{2} 	& \dots \\
		\BJ^{(2)}	& \frac{\BJ^{(1)}+\BJ^{(3)}}{2} & \frac{\BJ^{(4)}}{2}			& \frac{\BJ^{(1)}+\BJ^{(5)}}{2} 	& \frac{\BJ^{(2)}+\BJ^{(6)}}{2} 	& \dots \\
		\BJ^{(3)}	& \frac{\BJ^{(2)}+\BJ^{(4)}}{2} & \frac{\BJ^{(1)}+\BJ^{(5)}}{2}	& \frac{\BJ^{(6)}}{2} 				& \frac{\BJ^{(1)}+\BJ^{(7)}}{2} 	& \dots \\
		\BJ^{(4)}	& \frac{\BJ^{(3)}+\BJ^{(5)}}{2} & \frac{\BJ^{(2)}+\BJ^{(6)}}{2} & \frac{\BJ^{(1)}+\BJ^{(7)}}{2}		& \frac{\BJ^{(8)}}{2} 	& \dots \\
		\vdots & \vdots & \vdots & \vdots & \vdots & \ddots
	\end{pmatrix} + \bigoplus_{i=0}^\infty
	\begin{pmatrix}
		J_{11}^{(0)}-d_1\kappa_i	& J_{12}^{(0)} \\
		J_{21}^{(0)}	& J_{22}^{(0)}-d_2\kappa_i
	\end{pmatrix}
\end{equation}
does not have zero elements in general ($\BJ^{(k)}$ denotes 2x2 matrices), although analogous qualitative properties still hold. The appropriate conditions are constructed following the same idea:
\begin{equation}\label{Tconds}
	\begin{aligned}
		T1^R &:= b_{11} +s_{11} +b_{22}+s_{22} <0, \\
		T2^R &:= (b_{11}+s_{11})(b_{22}+s_{22}) - (b_{12}+s_{12})(b_{21}+s_{21}) >0, \\
		T3^R &:= (b_{11}+s_{11})d_2 +(b_{22}+s_{22})d_1 >0, \\
		T4^R &:= ((b_{11}+s_{11}) d_2 + (b_{22}+s_{22})d_1)^2-4d_1d_2((b_{11}+s_{11})(b_{22}+s_{22}) - (b_{12}+s_{12})(b_{21}+s_{21})) >0,
	\end{aligned}
\end{equation}
with $T1^L$-$T4^L$ of exactly the same form (i.e. $T1^R$-$T4^R$ with $s_{ij}\equiv 0$). The hypothesised conditions for pattern formation are of the same form as in Eqn. \eqref{lin_cond} and, as in the former case, were verified using both numerical studies of the model equations and by analysing the eigenvalues of truncated matrices from the spectral theory (results not shown).

\subsection{Computational approaches} \label{app.compu}

The spectrum of the truncated matrix \eqref{spec_matrix} is computed using the Matlab inbuilt function \textit{eig()}. When denoting $N$ as a constant representing the size of matrix ($(2N+2)\times (2N+2)$) our numerical results show that for $N>50$ the value of the largest real part of the eigenvalues does not significantly change; larger matrices contribute to the spectrum by eigenvalues with larger negative part as we discussed above. We choose $N=1000$. With constant $M$ representing the truncation in eigenmode expansion of $h(x)$, i.e. we approximate
$$h(x)\approx \sum_{k=0}^M Z^k y_k(x),\quad \text{where} |Z^k|\lessapprox \frac{s}{k}.$$
However, the length of numerical calculation does not significantly increase with larger $M$, so we set $M=N$.

The solutions of the evolution system \eqref{sys_RD_lin} are computed by Wolfram Mathematica 10 using \textit{NDSolve()} (via the method of lines for the temporal discretisation and finite differences for space) up to time $\tau=10^3$ or until the supremum norm of the solution exceeds $10^7$. The initial condition is random noise, uniformly distributed between $(10^{-2}, 10^2)$.

Both approaches to assess stability should yield the same result as they describe the same process. However, both methods are approximate and hence 
small differences might occur especially close to the border of the parameter regions due to different accuracy of the approximation (truncation of the matrix versus numerical discretisation when computing the evolution problem).




\end{document}